\documentclass[aps,twocolumn,prl,showpacs]{revtex4}
\usepackage{bm}
\usepackage{amsbsy}
\usepackage{graphicx}

\begin{document}

\newcommand{\bs}{\mathbf}
\newcommand{\tr}{\textrm}
\newcommand{\vp}{\varphi}
\newcommand{\e}{\epsilon}
\renewcommand{\d}{\delta}
\newcommand{\n}{{\hat n}}
\newcommand{\D}{{\tilde \Delta}}
\newcommand{\vD}{\bs \D}
\newcommand{\vt}{\bs \tau}
\newcommand{\h}{{\tilde h}}
\newcommand{\hb}{\hbar}
\newcommand{\s}{\sigma}
\newcommand{\vs}{{\bs \sigma}}
\newcommand{\lp}{\left}
\newcommand{\rp}{\right}
\newcommand{\ua}{\uparrow}
\newcommand{\ibid}{{\it ibid}}
\newcommand{\etl}{{\it et al.}}
\setlength\arraycolsep{1pt}

\title{Fully developed triplet proximity effect}

\author{V.~Braude and Yu.~V.~Nazarov}

\affiliation{Kavli Institute of Nanoscience, Delft University of Technology, 2628 CJ Delft,
   The Netherlands}

\date{\today}
\begin{abstract}
  We present a model for  fully developed triplet proximity effect in
 superconductor-ferromagnet heterostructures.
 Within the circuit-theory approximation, we evaluate 
 the Green's functions, the density of states, 
 and the Josephson current that depend essentially on
 the magnetization configuration. 
\end{abstract}
\maketitle
Research of 
heterostructures that combine superconducting (S) and ferromagnetic (F) elements gives insight to the problem of mutual influence
of superconductivity and ferromagnetism, 
allows realization of exotic superconducting states
such as the L-O-F-F state \cite{LOFF} and triplet ordering, \cite{review}
and promises
applications that utilize the spin degree of freedom. \cite{spintronics}
While this research has started more than three decades ago, 
\cite{old}  interest to the above topics has yet stipulated
recent important
developments, both theoretical and experimental.
Those concern Josephson $\pi$-junctions, \cite{pi,tripl2}
triplet superconductivity, \cite{tripl2,review,klapwijk}  and
Josephson spin valves. \cite{valve,absolute}

The superconducting proximity effect is characteristic for
most S/F heterostructures and is distinct from the effect
in non-magnetic S/N systems.
In the normal part of  a S/N structure,
the superconducting correlations persist at
distances of the order of normal-metal coherence length $\xi_N$.
This length scale can grow large at sufficiently small temperatures $T$. 
In a diffusive material, $\xi_N=\sqrt{\hbar D/2 \pi k_B T}$, $D$ being the 
diffusivity. 
In contrast to this,  superconducting correlations in a ferromagnet, 
where an exchange field $h$ is present, are quenched at much shorter scale 
$\xi_h=\sqrt{\hbar D/h}$. 
Hence one might conclude that the  proximity and Josephson effects 
are strongly suppressed 
in S/F heterostructures.
However, some experiments \cite{petrashov} seem to contradict  
 this statement, indicating  proximity correlations 
at much larger scale.
Though these experiments may be explained by interface effects, 
\cite{interface} they have motivated a proposal of an
 interesting mechanism for  long-range
proximity effect in ferromagnets  
\cite{tripl2,review}.
It was shown that inhomogeneity in the direction of exchange field 
generates superconducting
correlations of two electrons with the same spin, i.e. {\it triplet} correlations.
Such  triplet proximity effect (TPE) is not suppressed by an exchange field
and penetrates the ferromagnet at the scale of $\xi_N$.
Recently, a substantial Josephson current has been reported
for a fully polarized ferromagnet. \cite{klapwijk}
The experiment can only be explained by TPE.


An immediate problem is that the theoretical predictions
so far have been elaborated in assumption that  TPE 
is weak and can be treated perturbatively.
This makes it difficult to determine an unambiguous experimental signature 
of TPE to distinguish it from the conventional effect. 
Experimentally, the Josephson current due to TPE \cite{klapwijk}
does not seem to be smaller than that due to a fully
developed conventional proximity effect.

In this work, we address a fully developed TPE
that significantly changes the density of states (DOS) at the Fermi
level. We show that the DOS {\it increases}.
This is in contrast with
complete suppression of  DOS by a fully developed conventional 
proximity effect. Similar to the  conventional  effect,
the change in the DOS is restricted to the energy window $\simeq E_{Th}$,
the Thouless energy of the structure, provided $E_{Th} \ll \Delta$,
$\Delta$ being the energy gap in the superconductor. Therefore the corresponding
Josephson current is of the same order as for the conventional proximity
effect, though its magnitude  essentially depends
on the magnetization configuration in the structure. 
For magnets where both spin directions are present at the Fermi surface
(ferromagnetic metals), we find both 
 $\pi$-junctions \cite{pi,tripl2}
 (with negative supercurrents) and common $0$-junctions
(with positive supercurrents), depending on the magnetization directions.
For fully polarized magnetic materials (half-metals),
we find a continuous dependence of the equilibrium superconducting
phase difference on magnetization directions. 
Finite current may be induced thereby at fixed zero phase difference.

We concentrate on a S/F/S heterostructure
fabricated by deposition of two superconducting electrodes 
onto a conducting ferromagnetic film (Fig. 1a).\cite{klapwijk} 
We proceed with the so-called circuit theory  \cite{circuit}
that is a finite-element technique for semiclassical Green's function method 
\cite{quasicl}, which has been applied to S/F structures
in Ref.~\onlinecite{absolute}. Circuit theory is convenient since 
we aim at presenting an idealized TPE 
without consideration and subsequent optimization 
of concrete geometry of the structure. Besides, 
it allows simple analytical presentation of the results. 

We build a minimal circuit-theory model for  a ferromagnetic metal,
later adjusting it to a half-metal.
The left and right parts of the structure contain  regions where
superconducting and magnetic correlations meet (points $1$ and $3$
in Fig 1.a). Following \cite{absolute}, we represent each region
by a normal-metal node connected to a superconducting reservoir and 
a ferromagnetic insulating reservoir(FIR). 
The role of FIR
is to represent the exchange
field $h_k$ ($k=1,2,3$) induced in the node.
The middle of the structure (point $2$) is represented
as node $2$ connected to the nodes $1$,$3$, and 
another FIR. To enable TPE, we allow 
arbitrary magnetization directions of all  FIR's.
It is assumed for simplicity that all connectors are of tunnel nature.
The finite volume of each node and related decoherence between electrons
and holes is taken into account by introducing  "leakage" matrix currents
\cite{circuit} inversely proportional to the mean level spacing
$\delta_i$ in each node. This defines a circuit presented in Fig. 1b.

\begin{figure}
\begin{center}
\includegraphics[width=0.4\textwidth]{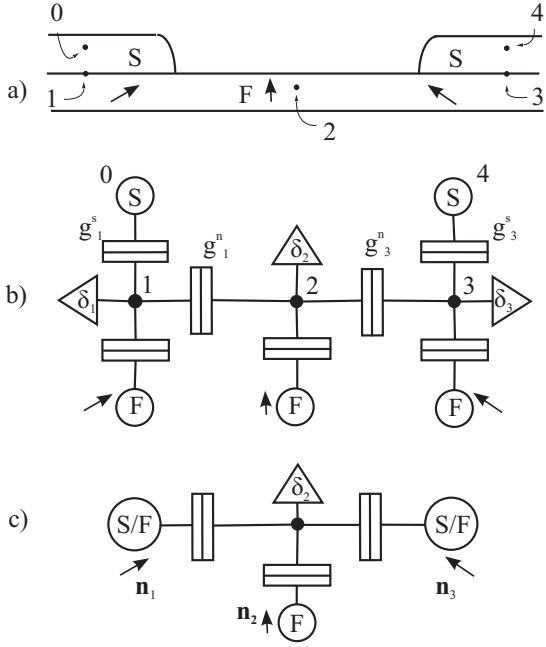}
\end{center}
\caption{Building up the circuit-theory model.
a)  S/F/S structure exhibiting  TPE \cite{klapwijk}.
b) The corresponding circuit consists of three nodes  representing
the parts $1-3$ of the structure.
c) The nodes $1$ and $3$ are replaced by effective
S/F reservoirs.}
\label{setup}
\end{figure}

The relevant variables are (retarded) 
Green's functions $G_{1,2,3}$ at the nodes. These Green's functions
are energy-dependent $4 \times 4$ matrices in the Nambu and spin space.
They are determined from the equations that express
conservation of the "matrix currents" at each node. 
These equations read $(k=1,3)$
\begin{eqnarray} 
\label{eq:cons}
 &&[ i\frac{ G_Q}{\delta_k}
h_k ({\bm{n}}_k \cdot \bm{\sigma}) \,\tau_3+\frac{g^s_{k}}{2} G^s_{k}-i\frac{ G_Q}{\d_k}\e \tau_3+\frac{g^n_{k}}{2} G_2,G_k]=0,
  \nonumber \\
 &&[i\frac{ G_Q}{\delta_2} h_2
  ({\bm{n}}_2 \cdot \bm{\sigma}) \tau_3-i\frac{ G_Q}{\d_2}\e \tau_3+
  \sum_{l=1,3} \frac{g^n_{l}}{2} G_l
 ,G_2]=0. 
  \quad
\end{eqnarray}

Here, $G_Q\equiv e^2/\pi\hb$ 
is the conductance quantum, $\bm{\tau}$ and $\bm{\sigma}$ 
are vectors composed of Pauli matrices 
in the Nambu and spin space, respectively.
The first term in these equations  is the
matrix current into the corresponding FIRs.  
The magnetization direction in  FIR $k$ is given
by a unit vector $\bm{n}_k$. 
 $G^s_k$ is a Green's function of
 superconducting reservoir $k$. 
We assume for simplicity that 
the relevant energies are
much smaller than the superconducting gap in the reservoirs, 
so we can use the energy-independent $G^s_k=\tau_1 \cos 
\vp_k+\tau_2 \sin \vp_k$, $\vp_k$ being the superconducting
phase.  Each superconducting reservoir $k$ is connected
to  node $k$ by a tunnel junction 
with  conductance $g^s_{k}$.
It would induce a proximity effect mini-gap 
$\Delta_k  =g^s_{k} \delta_{k}/2 G_Q$ 
in  node $k$ if no other connections were present.
The parameters $g^{n}_{1,3}$ give the conductances 
of connections between the normal nodes.

The two energy scales of the model are the typical
exchange field $h$ and the  Thouless energy 
$E_{Th} \simeq g^n \delta/G_Q$. The condition
$h \leq E_{Th}$ corresponds to a short S/F/S
structure with dimensions shorter or of the order
of $\xi_h$. In this case, the conventional proximity
effect overshadows TPE:
singlet anomalous components of $G$ 
in all nodes either exceed or are of the same order
as the triplet ones. Since we want to single out the
TPE, we turn to the opposite limit $E_{Th} \ll h$ of 
longer structures.

The best separation between the islands is achieved
in the limit $\d_2 \gg \d_{1,3}$, where
 the regions 
$1$ and $3$ adjacent to  the superconductors are
much longer than the middle part of the system.
We will see that in this limit the 
superconducting correlations
in the nodes $1$ and $3$ are between  electrons of 
opposite spin. As to the node $2$, the correlations 
are between electrons of the same spin only: there
 TPE is present in its purest form.

Under these conditions, the last term in the first Eq.~(\ref{eq:cons}) 
can be neglected and $G_{1,3}$ can be determined
separately from $G_2$. Owing to their big size, the
nodes $1$ and $3$ act as  effective reservoirs for the
node $2$ (Fig. 1c). 
The Green's functions at these nodes are determined
by the competition between the  corresponding superconducting reservoir
and FIR and read
\cite{absolute}  ($k=1,3;h_k>\Delta_k;\tau_{\pm} = 
\left[\tau_1 \pm i \tau_2\right]/2$):
\begin{equation} \label{eq:G1}
    G_{k}=\frac{ h_k \tau_3 -i(\Delta^*_k \tau_{+} + \Delta_k \tau_{-})
    (\bm{n}_k \cdot \bm{\sigma})}{\sqrt{h^2_k -|\Delta_k|^2}},
\end{equation}
where we assume $\epsilon \ll h_k, \Delta_k$. As seen from the 
structure of (\ref{eq:G1}), the effective S/F reservoirs supply
superconducting correlations that are different for opposite
spin directions. These correlations are  most pronounced if
$h_k \simeq \Delta_k$. We will see that this condition is  optimal
 for TPE. At this stage of research it is difficult to
immediately relate this condition to specifics of the structures in hand
such as geometry, film thickness, transparency of S/F interface etc.
Since the observations of \cite{klapwijk} suggest that
 TPE is close to the optimal one, we are convinced that this condition
is realizable. 

  Now the node $2$ is connected to  reservoirs only.
Its Green's function is determined by 
the balance of the matrix currents 
into these reservoirs. The two S/F reservoirs are connected 
in parallel, so that their net effect is additive and 
can be represented by
a matrix
$M = (g^{n}_1 G_1 + g^{n}_3 G_3)/(g^{n}_1+g^{n}_3) \equiv M_0\tau_3
-i(\bm{M}^* \tau_{+} + \bm{M} \tau_{-})\cdot \bm{\sigma}. 
$
It is important that the resulting $G_2$ splits into two {\it independent
blocks} corresponding to two spin projections on $\bm{n}_2$.
The separation into blocks allows us to treat half-metals
on equal footing with ferromagnetic metals. 
While for a ferromagnetic metal both blocks 
contribute to physical quantities, only a single block
does so in a half-metal. It is implied that in any case the S/F 
reservoirs 
support both spin directions, \cite{numeric} otherwise they would not be 
superconducting.

We write the block structure as follows: 
\begin{eqnarray} \label{eq:G2}
G_2 = S \left[ \begin{array}{cc} G_{\downarrow} & 0 \cr
0 & G_{\uparrow} 
\end{array}\right] S^{-1}; \\
S \equiv (1-\tau_3)/2-i\sigma_y
(\bm{n}_2 \cdot \bm{\sigma}) (1+\tau_3)/2. \nonumber
\end{eqnarray}
In these notations,
\begin{equation} \label{eq:Gup}
 G_{\uparrow} = \frac{1}{\sqrt{A_{\uparrow}}}\left[\begin{array}{cc}
M_0  - i \epsilon/E_{Th} &  \quad -i( M^*_x - i M^*_y) \cr
-i (M_x + i M_y) & \quad -M_0 + i \epsilon/ E_{Th}
\end{array}\right],   
\end{equation}
where $ A_{\uparrow} \equiv (M_0 -i\epsilon/E_{Th})^2 - |M_x +i M_y|^2$ 
and the $z$ axis 
is chosen in the direction of $\bm{n}_2$. The result 
for $G_{\downarrow},A_{\downarrow}$ 
is obtained by replacing $M_x\pm i M_y$ 
with $M_x\mp i M_y$. The advantage of such notations is
that the block structure is made explicit. 
The non-diagonal 
elements of $G_{\uparrow}$ correspond to triplet  anomalous
averages $\langle\psi_{\uparrow}\psi_{\uparrow} \rangle$,
$\langle\psi^{\dagger}_{\uparrow}\psi^{\dagger}_{\uparrow} \rangle$,
whereas the singlet-pairing averages $\langle\psi_{\uparrow}\psi_{\downarrow} 
\rangle$,
$\langle\psi^{\dagger}_{\uparrow}\psi^{\dagger}_{\downarrow} \rangle$
vanish.
This manifests a pure TPE. The fully developed effect is characterized 
by 
$h_k \gtrsim |\Delta_k|$, in which case the non-diagonal and diagonal elements
of $G_{\uparrow,\downarrow}$ are of the same order of magnitude.

The triplet anomalous averages $\langle\psi_{\uparrow}\psi_{\uparrow} 
\rangle$
($\langle\psi_{\downarrow}\psi_{\downarrow} 
\rangle$)
 acquire a phase
factor $e^{ i\chi}$ ($e^{ -i\chi}$) upon  rotation by an angle $\chi$ 
about the $z$ axis. This leads to an interesting interplay
between the superconducting phase difference $\phi_3-\phi_1 \equiv \phi$
and the relative longitude angle between $\bm{n}_1$ and $\bm{n}_3$, 
$\chi_3-\chi_1 \equiv \chi$.
To see this explicitly, we express the magnetization direction
vectors $\bm{n}_{1,3}$ in the spherical coordinates ($\chi,\theta$ standing
for the longitude and latitude respectively). 
In these notations, 
\begin{equation}
M_x+ i M_y = a_1 e^{i(\phi_1 +\chi_1)}   + 
a_3 e^{i(\phi_3 +\chi_3)}, 
\end{equation}
where 
$a_k \equiv g^n_k|\Delta_k|\sin\theta_k/(g^n_1+g^n_3)\sqrt{h^2_k -|\Delta_k|^2}$,
so that the superconducting phase and the longitude always
come together.

\begin{figure}
\begin{center}
\makebox[ \textwidth][l]{
\hspace{0.02\textwidth}
\includegraphics[width=0.28\textwidth]{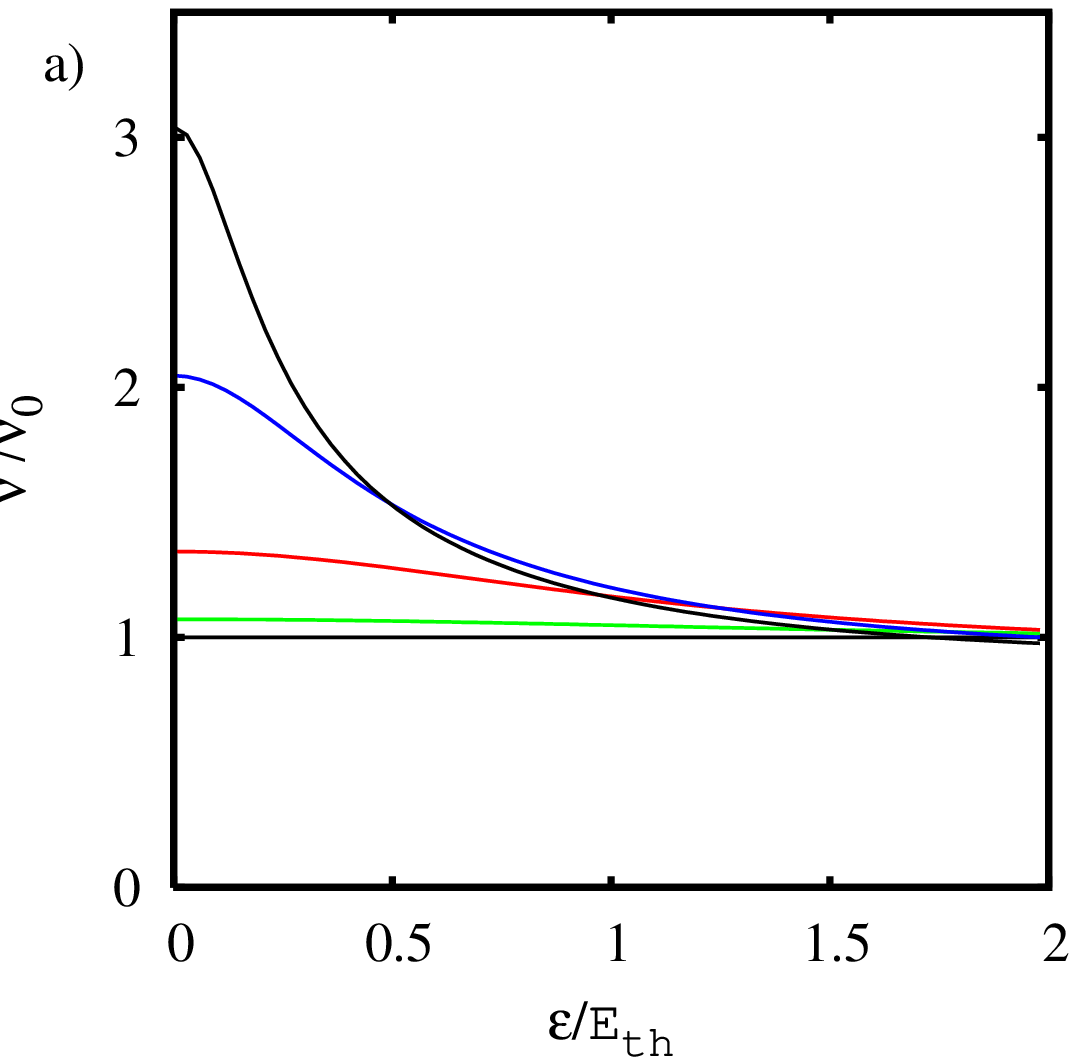}
\hspace{-0.06\textwidth}
\includegraphics[width=0.28\textwidth]{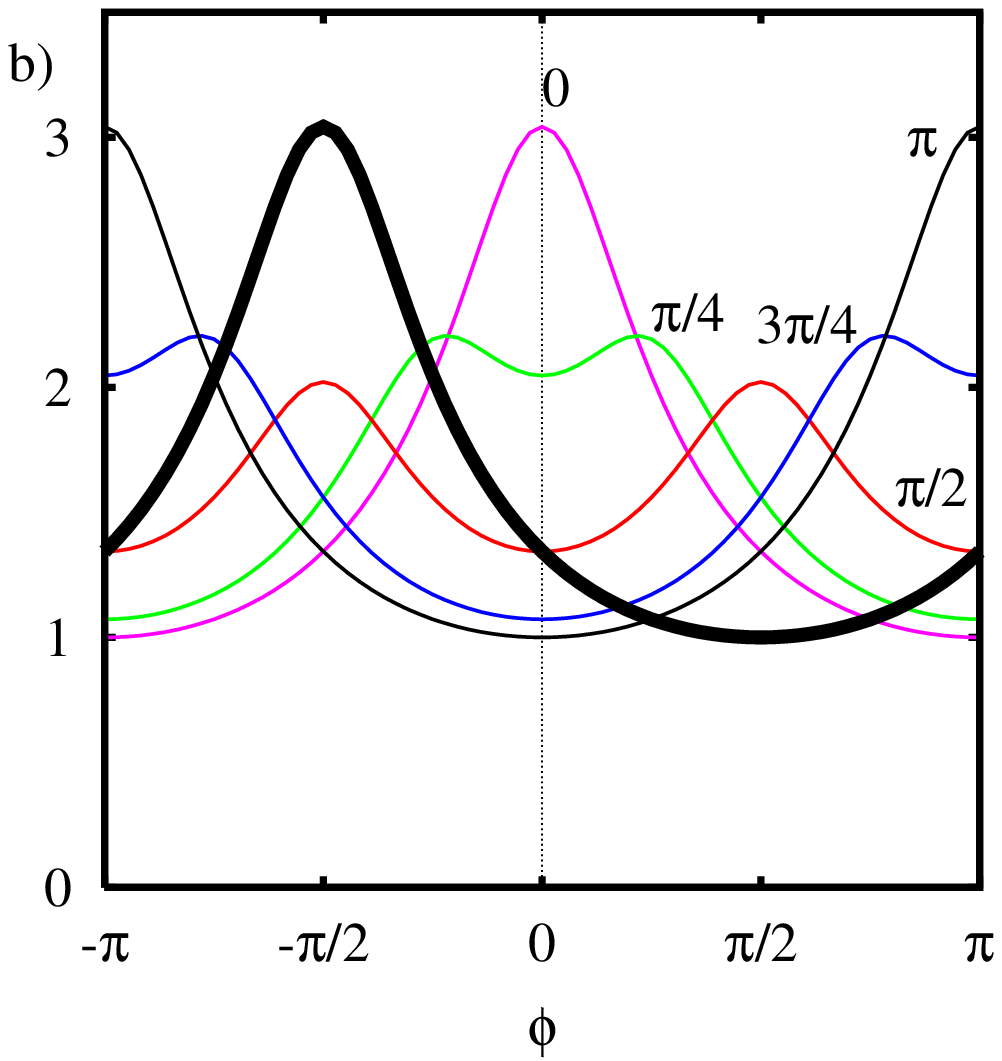}
}
\end{center}
\caption{ The DOS in the central 
node a) versus the energy for $\phi=0$ and $\chi=0$ where $\theta_{1}=\theta_3$ increases from $0$ for the lower 
curve to  $\pi/2$ for the upper one
 with a step $\pi/8$. 
b) The total DOS for a ferromagnetic metal at zero energy versus $\phi$ 
for $\theta_1=\theta_3=\pi/2$ where  $\chi$ increases from $0$ to $\pi$ with a step $\pi/4$.
Upon increasing $\chi$ a single peak at $\phi=0$ splits
in  two. The thick curve corresponds to  a half-metal with $\chi=\pi/2$.
For both graphs,  $h_{1,3}/\Delta_{1,3}=18/17$. }
\label{fig:nu-en}
\end{figure}

The DOS is determined from $G_2$
and generally is different for the opposite spin directions,
\begin{equation}
\nu^{\uparrow,\downarrow}(\varepsilon) = \nu^{\uparrow,\downarrow}_0
{\rm Re} 
\left( 1- \frac {a_1^2+a_3^2 + 2a_1 a_3 \cos(\phi \pm \chi)}
{ (M_0-i\e/E_{Th})^2}
\right)^{-\frac{1}{2}},
\label{eq:nuupdown}
\end{equation}
 where $\nu^{\uparrow,\downarrow}_0$ is the normal-state DOS. 
 As shown in Fig.~\ref{fig:nu-en}(a), 
$\nu$ always exceeds the normal-state value at $\varepsilon \ll E_{Th}$.
This is in contrast with a suppression of $\nu$ manifesting
the conventional proximity effect. Such an enhanced DOS 
is therefore a signature of TPE. The peak at small
energies is followed by a wider dip at energies of the order of
the Thouless energy $E_{Th}$ so that the total number of states
 remains
unchanged [Fig. \ref{fig:nu-en}(a)]. The dependence of 
$\nu$ on the superconducting
phase/longitude is also characteristic of TPE. 
As seen from Eq.~(\ref{eq:nuupdown}), for a given spin direction the
effect of superconducting phase can be always compensated by a
rotation of one of the magnetization directions $\bm{n}_{1,3}$
 about $\bm{n}_2$. 
While for a ferromagnetic metal the total DOS $\nu=\nu^{\uparrow}+
\nu^{\downarrow}$ is 
an even function of $\phi$ and $\chi$, [Fig. \ref{fig:nu-en}(b)] 
this is generally not so 
for a half-metal. In the latter case, for a symmetric setup $a_1=a_3$, 
the DOS
 $\nu(0)$  can be modulated by either phase from $1$ to its maximum
value $1/\sqrt{1-4a_1^2/M_0^2}$.

\begin{figure}[b]
\begin{center}
\makebox[ \textwidth][l]{
\hspace{0.04\textwidth}
\includegraphics[width=0.28\textwidth]{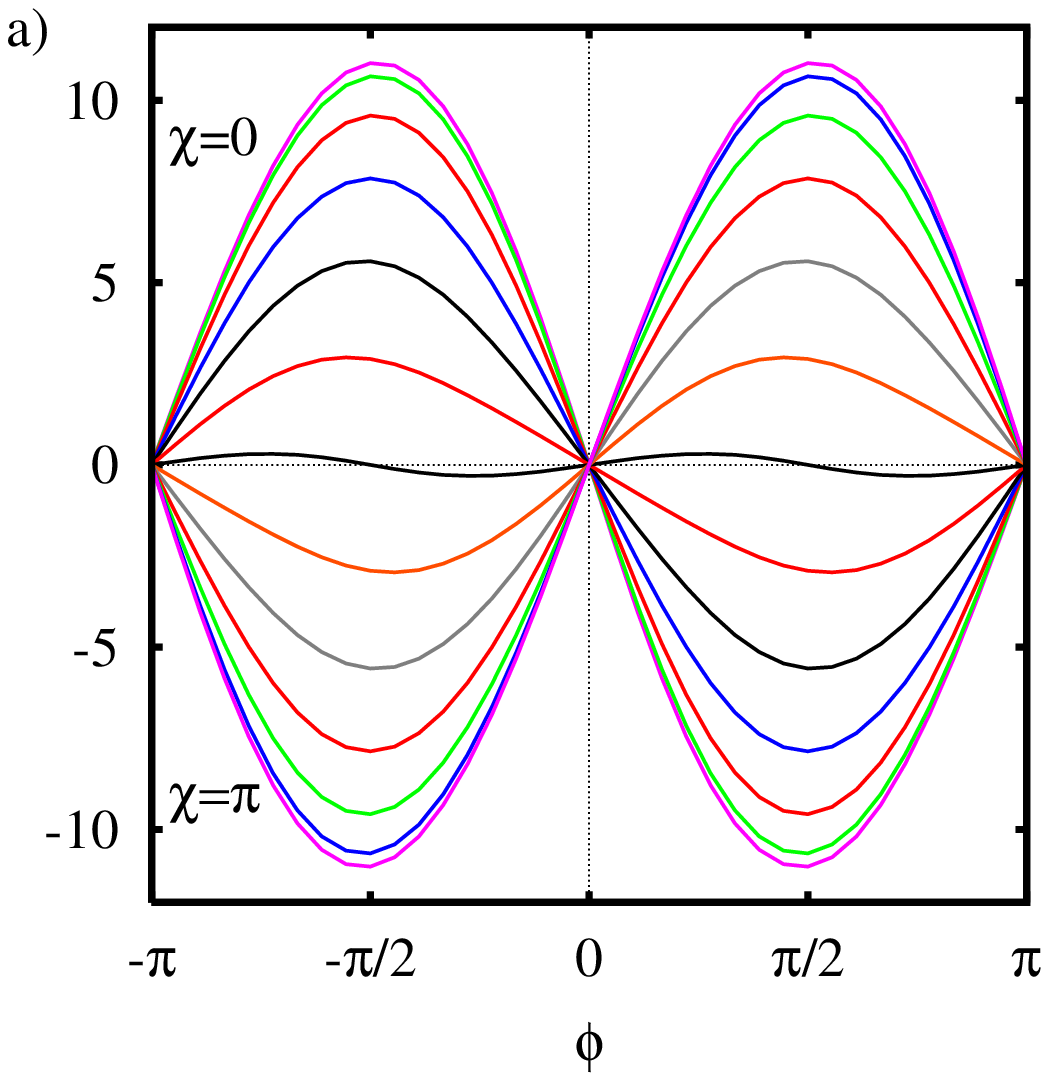}
\hspace{-0.06\textwidth}
\includegraphics[width=0.28\textwidth]{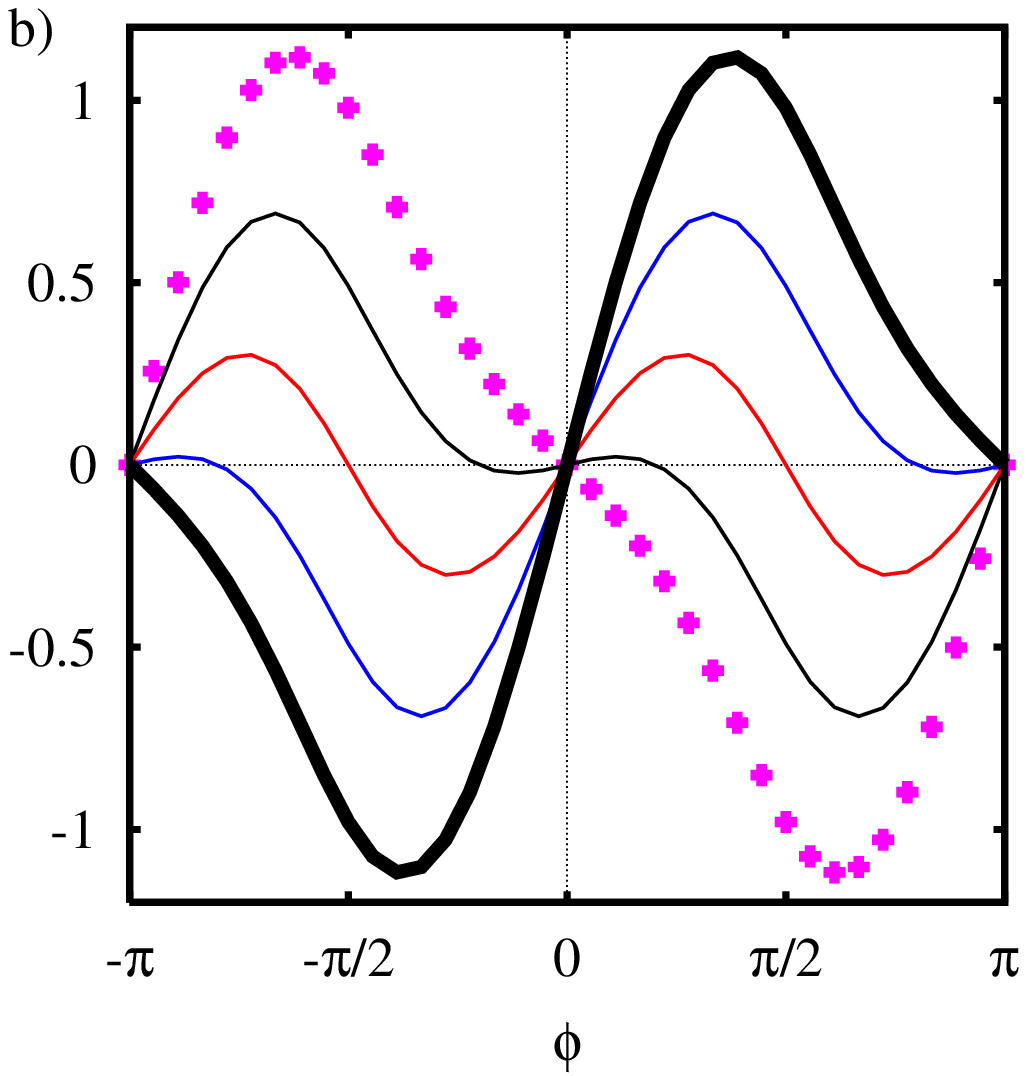}
}
\end{center}
\caption{The Josephson current for a ferromagnetic metal
versus $\phi$ for different $\chi$:
a) varying from $0$ to $\pi$ with a step $\pi/12$. 
The first harmonic
dominates the current except the vicinity of $\chi=\pi/2$ 
where it changes the sign. 
b) The vicinity of ``$0$''-``$\pi$'' transition.
Here $\chi$ varies from $\pi/2-\pi/36$ (``$\pi$''-state, dotted line) 
through
$\pi/2-\pi/72, \pi/2, \pi/2+\pi/72$ (bistable, thin lines) to
$\pi/2+\pi/36$ (``$0$''-state, thick line).}
\label{fig:J1}
\end{figure}

\begin{figure}[b] 
\begin{center}
\includegraphics[angle=0, width=0.35\textwidth]{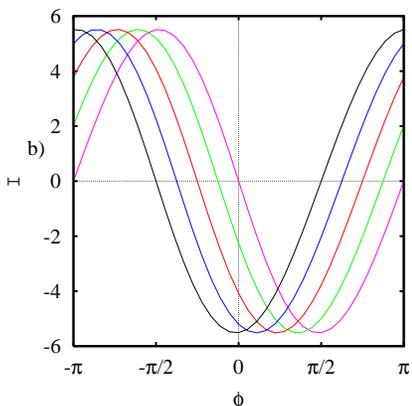}
\end{center}
\caption{Tne Josephson current for a half-metal versus $\phi$
for $\chi$ varying from $0$ to $\pi/2$ with a step $\pi/8$.
 The $I-\phi$ curves shift to the left with increasing $\chi$.}
\label{fig:J2}
\end{figure}

In the  model under consideration, 
the Josephson current is 
given by the Keldysh component of the matrix current through
either junction $g^n_{1,3}$ integrated over energy,
\begin{equation}
  I=-\frac{g^n_{1}}{4 e}\int_{-\infty}^\infty d\e \,
 \tr{Re}\,  \tr{Tr}\,\tau_3 [G_1,G_2] \tanh \frac{\e}{2T}.
 \label{current-general}
\end{equation}
The block structure of  $G_2$ gives rise to 
two contributions to the current that correspond  to
 opposite spin directions. The integral in 
Eq.~(\ref{current-general}) is logarithmic, converging
at energies $E_{Th} \ll \epsilon \ll \Delta_{cut}$,
$\Delta_{cut} \simeq {\rm min}(\Delta,h,h-\Delta)$. 
We assume $T\ll E_{Th}$.
With  logarithmic accuracy,
\begin{equation}
   I_{\uparrow,\downarrow}=-\frac{2 a_1 a_3 (g^n_1+g^n_3)}{e}
   \sin(\phi \pm \chi)
    E_{Th} \ln\left( 
    \frac{\Delta_{cut}}{E_{Th}}
    \right) .
 \label{j-current}
 \end{equation}
 Both the scale of the current ($I \simeq g E_{Th}/e$)
 and the logarithmic structure 
 are similar to the common proximity
 effect. The difference is the dependence of the
 current on the magnetization directions.
 For a ferromagnetic metal, the total current $I_{\uparrow} +
 I_{\downarrow}$ is odd in $\phi$.
 Interestingly, the sign of the Josephson current 
 is opposite to that 
 in a common Josephson contact provided $|\chi|<\pi/2$. 
 This signals a $\pi$-junction \cite{pi, tripl2, pi2},
 which can be changed to a common $0$-junction by changing $\chi$.
 Since the accuracy of the logarithmic approximation
 is always questionable, we evaluate the integral 
 numerically assuming  $\Delta_{cut}/E_{Th} =50$.
 The resulting dependence of the current on both phases 
 is not precisely  harmonic, though close to it. (Figs. 
 \ref{fig:J1},\ref{fig:J2}). As seen in Fig. 
 \ref{fig:J1}(b), the second harmonic of the current 
 becomes dominant in the vicinity of $\chi =\pi/2$. 
 This implies that the transition between "$\pi$" and "0" states 
 follows a scenario of Ref. \onlinecite{Chelkachev} where 
 both states are stable in this vicinity. 

For a half-metal, the situation is very different. In this
case, the superconducting phase corresponding to the energy minimum
just follows $\chi$ and changes continuously instead of "jumping"
between the values of $0$ and $\pi$. There is a finite
supercurrent at zero phase difference. Therefore,  rotation
of the magnetizations $\bm{n}_{1,3}$ about $\bm{n}_2$ is 
equivalent to the
effect of extra magnetic flux $\chi \Phi_0/2\pi$ enclosed 
in a  large loop that includes the junction. In our opinion, 
this facilitates
an unambiguous experimental verification of  TPE.

In both cases, not only the supercurrent is zero at the energetically 
favorable phase difference, but also  TPE is reduced, vanishing 
completely for an ideally symmetric setup. This is because  TPE 
increases the electron energy in the central node, as seen from the fact
that the DOS is enhanced at low energies.

In conclusion, 
we have proposed a simple model for fully developed 
triplet proximity effect in  S/F/S 
structures. In contrast to the common proximity effect,
 TPE enhances the DOS at low energies.
The Josephson current exhibits a peculiar dependence
on the magnetization
configuration that is essentially different for 
a ferromagnetic  metal and a half-metal. 
Those are  signatures of  TPE to be
observed experimentally.

We appreciate useful discussions with Ya. M. Blanter, G.~E.~W.~Bauer,
 T.~M.~Klapwijk, R.~S.~Keizer, and E.~B.~Sonin and valuable communications
 with Y.~Asano and A.~F.~Volkov. The work was supported by 
 EC NMP2 - CT2003-505587 "SFINX" .

\end{document}